%% file: 0-main.tex
\documentclass{article}
\usepackage{ijcai26}

\usepackage{times}
\usepackage{soul}
\usepackage{url}
\usepackage[hidelinks]{hyperref}
\usepackage[utf8]{inputenc}
\usepackage[small]{caption}
\usepackage{graphicx}
\usepackage{amsmath}
\usepackage{amsthm}
\usepackage{booktabs}
\usepackage{algorithm}
\usepackage{algorithmic}
\usepackage[switch]{lineno}
\usepackage{times}
\usepackage{soul}
\usepackage{multirow}
\usepackage{url}
\usepackage[hidelinks]{hyperref}
\usepackage[utf8]{inputenc}
\usepackage[font=small]{caption}
\usepackage{graphicx}
\usepackage{amsmath}
\usepackage{amsthm}
\usepackage{booktabs}
\usepackage{algorithm}
\usepackage{algorithmic}
\usepackage[switch]{lineno}
\usepackage{bm}
\usepackage{array}
\usepackage{amsfonts}
\usepackage{color}
\usepackage{enumitem}
\usepackage{amsthm}
\usepackage{caption}

\usepackage{tikz}
\usepackage[disable]{todonotes}
\usetikzlibrary{fadings}
\usetikzlibrary{mindmap,trees}
\usetikzlibrary{arrows,automata,shapes,positioning,shadows,trees}
\usetikzlibrary{shapes,snakes,shadows}
\usetikzlibrary{shapes.arrows}
\usetikzlibrary{calc,shapes, positioning}
\usetikzlibrary{decorations.text}
\usetikzlibrary{matrix,chains,positioning,decorations.pathreplacing,arrows}
\usetikzlibrary{bayesnet}
\usepackage[edges]{forest}
\usetikzlibrary{shadows.blur}
\usetikzlibrary{shapes.geometric}
\tikzstyle{edge}=[-latex',draw=black!90,shorten <=1pt,shorten >=1pt]
\tikzstyle{redge}=[latex'-,draw=black!90,shorten <=1pt,shorten >=1pt]
\tikzstyle{dedge}=[latex'-latex',draw=black!90,shorten <=1pt,shorten >=1pt]

\tikzstyle{block}=[draw, text width=5em,align=center,shape=rectangle, rounded corners, , align=center]
\tikzstyle{nobox}=[align=center]
\definecolor{emb}{RGB}{209,228,252}
\definecolor{hidden-blue}{RGB}{194,232,247}
\definecolor{hidden-orange}{RGB}{224,224,224}
\definecolor{hidden-yellow}{RGB}{242,244,193}
\definecolor{output-purple}{RGB}{219,203,231}
\definecolor{output-green}{RGB}{204,231,207}
\definecolor{output-black}{RGB}{0,0,0}
\definecolor{output-white}{RGB}{255,255,255}
\definecolor{myorange}{RGB}{255,208,153}

\definecolor{mygreen}{RGB}{166,207,152} % 淡色
\definecolor{hiddendraw-green}{RGB}{85,124,85}
\definecolor{myred}{RGB}{207,153,150}
\definecolor{hiddendraw-red}{RGB}{178,87,81}
\definecolor{mypurple}{RGB}{176,163,196}
\definecolor{hiddendraw-purple}{RGB}{124,101,158}
\definecolor{myblue}{RGB}{47, 110, 186}

\tikzstyle{emb-purple}=[
    rectangle,
    draw=output-purple!50!purple,
    fill=output-purple,
    text opacity=1,
    minimum height=1.5em,
    minimum width=1.5em,
    inner sep=0pt,
    align=center,
    fill opacity=.5,
    ]
\tikzstyle{emb-blue}=[
    rectangle,
    draw=emb!50!blue,
    fill=emb,
    text opacity=1,
    minimum height=1.5em,
    minimum width=1.5em,
    inner sep=0pt,
    align=center,
    fill opacity=.5,
]

\title{Trustworthy Intelligent Education: A Systematic Perspective on \\Progress, Challenges, and Future Directions}

\author{
Xiaoshan Yu$^1$ \and
Shangshang Yang$^2$\and
Ziwen Wang$^2$\and\\
Haiping Ma$^3$\and 
Xingyi Zhang$^2$
\affiliations
$^1$School of Artificaial Intelligence, Anhui University\\
$^2$School of Computer Science and Technology, Anhui University\\
$^3$Institutes of Physical Science and Information Technology, Anhui University\\
\emails
\{yxsleo, yangshang0308, wzw12sir, xyzhanghust\}@gmail.com, hpma@ahu.edu.cn
}

\newtheorem*{Problem Definition}{Problem Definition}

\begin{document}

\maketitle
\input{1-abstract}

\input{2-introduction}

\input{3-intel_edu_tasks}

\input{4-trustworthy_IE}

\input{5-future_directions}

\input{6-conclusion}
% \vspace{-2mm}

% \newpage

% \newpage
\bibliographystyle{named}
\bibliography{reference}

\end{document}

%% file: 1-abstract.tex
\begin{abstract}

In recent years, trustworthiness has garnered increasing attention and exploration in the field of intelligent education, due to the inherent sensitivity of educational scenarios, such as involving minors and vulnerable groups, highly personalized learning data, and high-stakes educational outcomes. However, existing research either focuses on task-specific trustworthy methods without a holistic view of trustworthy intelligent education, or provides survey-level discussions that remain high-level and fragmented, lacking a clear and systematic categorization. To address these limitations, in this paper, we present a systematic and structured review of trustworthy intelligent education. Specifically, We first organize intelligent education into five representative task categories: learner ability assessment, learning resource recommendation, learning analytics, educational content understanding, and instructional assistance. Building on this task landscape, we review existing studies from five trustworthiness perspectives, including safety and privacy, robustness, fairness, explainability, and sustainability, and summarize and categorize the research methodologies and solution strategies therein. Finally, we summarize key challenges and discuss future research directions. This survey aims to provide a coherent reference framework and facilitate a clearer understanding of trustworthiness in intelligent education.

\end{abstract}

%% file: 2-introduction.tex
\section{Introduction}

With the rapid advancement of deep learning and data-driven technologies, intelligent education~\cite{AIED} has emerged as a significant paradigm within modern educational technology. Unlike traditional education modes that primarily rely on static curricula and uniform teaching approaches, intelligent education emphasizes adaptability and personalization. It aims to dynamically model students' knowledge levels, behavioral performance, and learning needs while delivering tailored learning resources and instructional support. Moreover, this paradigm has been widely adopted across various online learning platforms and digital educational infrastructure, reshaping how teaching and learning are conducted.

As intelligent education systems increasingly participate in high-impact educational decisions, trustworthiness has become an unavoidable concern~\cite{smuha2019eu}. This is because educational scenarios are inherently sensitive: learners are typically minors, learning data captures fine-grained behaviors, and system outputs can tangibly influence learners' educational trajectories. Therefore, untrustworthy behaviors, including privacy leakage, unstable predictions, biased decisions, and opaque recommendations, may lead to not only technical failures but also pedagogical and social risks.

In recent years, growing research has begun exploring trust-related issues in intelligent education, with topics including privacy cognitive diagnosis~\cite{FedCD}, robust learning modeling~\cite{AdaRD}, and explainable educational assessment~\cite{LACAT}. Although these efforts advance trustworthy techniques from diverse perspectives, they remain scattered across individual tasks. Meanwhile, existing surveys~\cite{smuha2020trustworthy,anagnostopoulou2024trustworthy,li2025trustworthy} that attempt to explore the intersection between trustworthy AI and education are often high-level or fragmented, and tend to adopt inconsistent task~classification.

To this end, this paper aims to address these limitations by providing a systematic and structured survey of trustworthy intelligent education, as shown in Figure~1. Specifically, we first outline a clear landscape of intelligent education by identifying five representative task categories: learner ability assessment, learning resource recommendation, learning analytics, educational content understanding, and instructional assistance. Building upon this task landscape, we conduct a comprehensive review of existing studies from five core trustworthiness perspectives: safety and privacy, robustness, fairness, explainability, and sustainability. Through this dual-level organization, we reveal how different trustworthiness concerns arise, are addressed, and interact across intelligent education tasks. Furthermore, we summarize key open challenges and discuss promising future research directions, aiming to provide a coherent conceptual framework and a practical roadmap for advancing trustworthy intelligent education.

\tikzstyle{leafs1}=[draw=hiddendraw-green,
    rounded corners,minimum height=1em,
    fill=mygreen!40,text opacity=1, align=center,
    fill opacity=.3,  
    text=black,
    font=\scriptsize,
    inner xsep=3pt,
    inner ysep=1pt,
    ]
\tikzstyle{leafs2}=[draw=hiddendraw-red,
    rounded corners,minimum height=1em,
    fill=myred!40,text opacity=1, align=center,
    fill opacity=.3,  
    text=black,
    font=\scriptsize,
    inner xsep=3pt,
    inner ysep=1pt,
    ]
\tikzstyle{leafs3}=[draw=hiddendraw-purple,
    rounded corners,minimum height=1em,
    fill=mypurple!40,text opacity=1, align=center,
    fill opacity=.3,  
    text=black,
    font=\scriptsize,
    inner xsep=3pt,
    inner ysep=1pt,
    ]
\tikzstyle{middles1}=[draw=hiddendraw-green,
    rounded corners,minimum height=1em,
    fill=mygreen!50,
    text opacity=1, 
    align=center,
    fill opacity=.5,  
    text=black,
    font=\scriptsize,
    inner xsep=3pt,
    inner ysep=1pt,
    ]
\tikzstyle{middles2}=[draw=hiddendraw-red,
    rounded corners,minimum height=1em,
    fill=myred!40,
    text opacity=1, 
    align=center,
    fill opacity=.5,  
    text=black,
    font=\scriptsize,
    inner xsep=3pt,
    inner ysep=1pt,
    ]
\tikzstyle{middles3}=[draw=hiddendraw-purple,
    rounded corners,minimum height=1em,
    fill=mypurple!40,
    text opacity=1, 
    align=center,
    fill opacity=.5,  
    text=black,
    font=\scriptsize,
    inner xsep=3pt,
    inner ysep=1pt,
    ]
\tikzstyle{roots}=[draw=output-black,
    % rounded corners,
    minimum height=1em,
    fill=myblue!60,
    text opacity=1, 
    align=center,
    fill opacity=.5,  
    text=black,
    font=\scriptsize,
    inner xsep=3pt,
    inner ysep=1pt,
    ]
\newcommand{\twtwo}{8.7em}
\newcommand{\twthree}{9.3em}
\newcommand{\twfour}{25em}
\newcommand{\twfive}{13.9em}
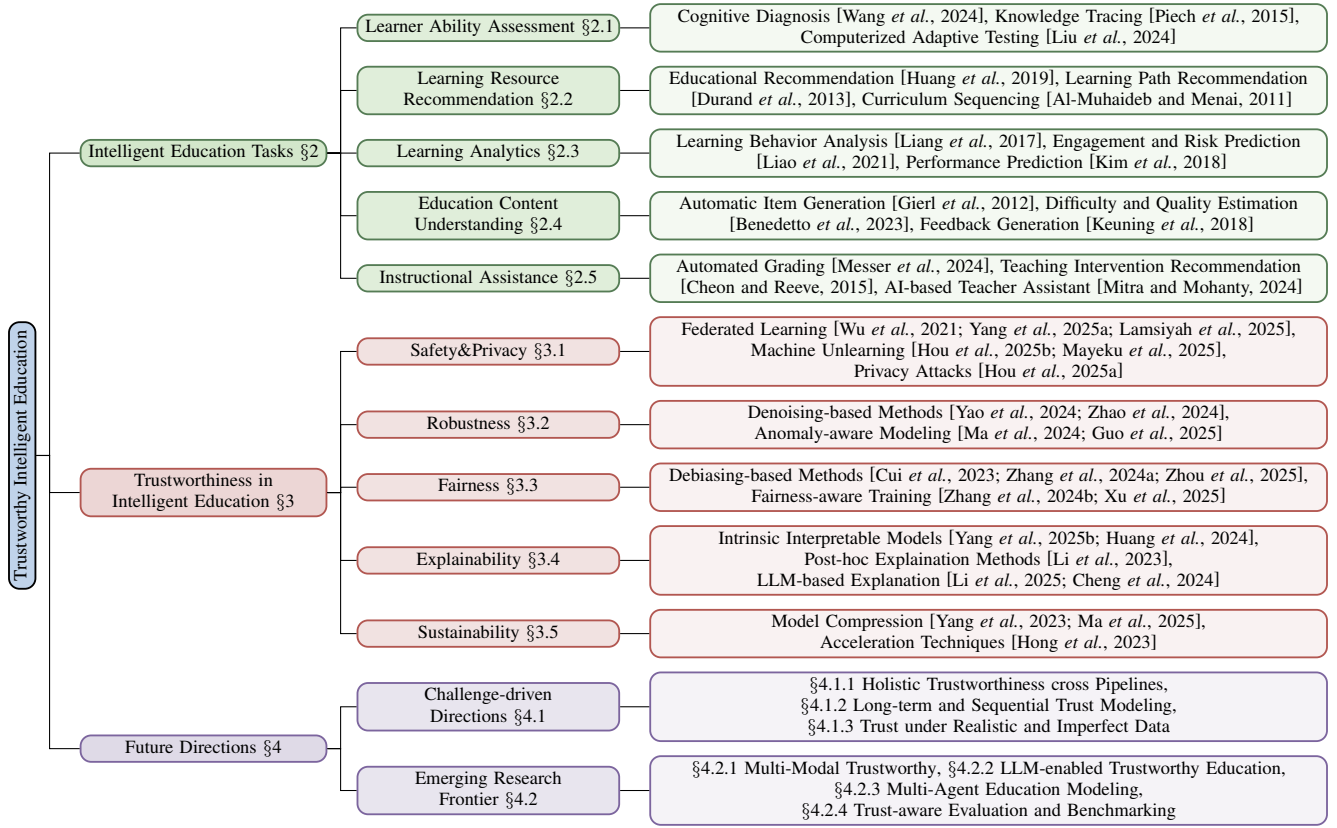
\begin{figure*}[ht]
\centering
\begin{forest}
  for tree={
    forked edges,
    grow=east,
    reversed=true,
    anchor=base west,
    parent anchor=east,
    child anchor=west,
    base=roots,
    font=\scriptsize,
    rectangle,
    line width=0.7pt,
    draw=output-black,
    rounded corners,
    align=center,
    minimum width=2em,
    s sep=5pt,
    inner xsep=3pt,
    inner ysep=1pt,
  },
  where level=1{text width=4.5em}{},
  where level=2{text width=6em, font=\scriptsize}{},
  where level=3{font=\scriptsize}{},
  where level=4{font=\scriptsize}{},
  where level=5{font=\scriptsize}{},
  [Trustworthy Intelligent Education, roots,
  % fill=myblue!40,
  rotate=90,
  anchor=north,
  edge=output-black
    [Intelligent Education Tasks~$\S$2, middles1, align=center, 
    fill opacity=1., edge=output-black,text width=\twtwo
        [Learner Ability Assessment~$\S$2.1, 
        middles1, align=center, 
        fill opacity=.7, 
        text width=\twthree, edge=output-black
            [Cognitive~Diagnosis~\cite{wang2024survey}{, }Knowledge~Tracing~\cite{DKT}{, }\\Computerized~Adaptive Testing~\cite{CAT_Survey}, leafs1, align=center, text width=\twfour, edge=output-black]
        ]
        [Learning Resource \\Recommendation~$\S$2.2, middles1, fill opacity=.7, align=center, text width=\twthree, edge=output-black
            [Educational Recommendation~\cite{Exercise_Rec_1}{, }Learning~Path Recommendation\\~\cite{LPR_Graph}{, }Curriculum Sequencing~\cite{Curriculum_Seq_Evo}, leafs1, align=center, text width=\twfour, edge=output-black]        
        ]
        [Learning Analytics~$\S$2.3, middles1, fill opacity=.7, align=center, edge=output-black, text width=\twthree
            [Learning Behavior Analysis~\cite{Behavior_Analysis}{, }Engagement and Risk Prediction\\~\cite{Engagement_Pre}{, }Performance Prediction~\cite{Performance_Pre}, leafs1, align=center, text width=\twfour, edge=output-black]
        ]
        [Education Content \\Understanding~$\S$2.4, middles1, fill opacity=.7, align=center, edge=output-black, text width=\twthree
            [Automatic Item Generation~\cite{Auto_Item_Gen}{, }Difficulty and Quality Estimation\\~\cite{Diff_Estimation_Survey}{, }Feedback Generation~\cite{Feedback_Gen}, leafs1, align=center, text width=\twfour, edge=output-black]
        ]
        [Instructional Assistance~$\S$2.5, middles1, fill opacity=.7, align=center, edge=output-black, text width=\twthree
            [Automated Grading~\cite{Automated_Grading}{, }Teaching Intervention Recommendation\\~\cite{Teaching_Intervention}{, }AI-based Teacher Assistant~\cite{Teaching_Assistant}, leafs1, align=center, text width=\twfour, edge=output-black]
        ]
    ] 
    [Trustworthiness in \\Intelligent Education~$\S$3, middles2, fill opacity=1., align=center, edge=output-black,text width=\twtwo
        [Safety\&Privacy~$\S$3.1, middles2, fill opacity=.7, align=center, text width=\twthree, edge=output-black
            [Federated Learning~\cite{FDKT,FedCD,FLSD}{, }\\Machine Unlearning~\cite{PrivacyCD,MU4RAAI}{,}\\Privacy Attacks~\cite{P-MIA}, leafs2, align=center, text width=\twfour, edge=output-black]
        ]
        [Robustness~$\S$3.2, middles2, fill opacity=.7, align=center, text width=\twthree, edge=output-black
            [Denoising-based Methods~\cite{AdaRD,DiffCog}{, }\\Anomaly-aware Modeling~\cite{HD-KT,RobustKT}, leafs2, align=center, text width=\twfour, edge=output-black]      
        ]
        [Fairness~$\S$3.3, middles2, fill opacity=.7, align=center, edge=output-black, text width=\twthree
            [Debiasing-based Methods~\cite{CORE,PSCRF,DisKT}{, }\\Fairness-aware~Training~\cite{FairCD,FairWISA}, leafs2, align=center, text width=\twfour, edge=output-black]
        ]
        [Explainability~$\S$3.4, middles2, fill opacity=.7, align=center, edge=output-black, text width=\twthree
            [Intrinsic Interpretable Models~\cite{KAN2CD,XKT}{, }\\Post-hoc Explaination Methods~\cite{GCE}{, }\\LLM-based Explanation~\cite{EFKT,LACAT}, leafs2, align=center, text width=\twfour, edge=output-black]
        ]
        [Sustainability~$\S$3.5, middles2, fill opacity=.7, align=center, edge=output-black, text width=\twthree
            [Model Compression~\cite{ENAS-KT,HashCAT}{, }\\Acceleration~Techniques~\cite{SECAT}, leafs2, align=center, text width=\twfour, edge=output-black]
        ]  
    ]
    [Future Directions~$\S$4, middles3, fill opacity=1., align=center, edge=output-black,text width=\twtwo
        [Challenge-driven \\Directions~$\S$4.1, middles3, fill opacity=.7, align=center, text width=\twthree, edge=output-black
            [$\S$4.1.1~Holistic Trustworthiness cross Pipelines{,}\\$\S$4.1.2~Long-term and Sequential Trust Modeling{,}\\$\S$4.1.3~Trust under Realistic and Imperfect Data, leafs3, align=center, text width=\twfour, edge=output-black]
        ]
        [Emerging Research \\Frontier~$\S$4.2, middles3, fill opacity=.7, align=center, text width=\twthree, edge=output-black
            [$\S$4.2.1~Multi-Modal Trustworthy{,} $\S$4.2.2~LLM-enabled Trustworthy Education{,}\\$\S$4.2.3~Multi-Agent Education Modeling{,}\\ $\S$4.2.4~Trust-aware Evaluation and Benchmarking, leafs3, align=center, text width=\twfour, edge=output-black]
        ]
    ]  
]
\end{forest}% Ensure all braces are properly matched above this line
\caption{A Taxonomy of Trustworthy Intelligent Education.}
\label{fig:taxonomy}
\end{figure*}

In summary, this survey makes the following contributions:

\begin{itemize}[leftmargin=1.5em]

    \item[\small$\bullet$] We provide a structured view of intelligent education by organizing its core functionalities into five key task types.

    \item[\small$\bullet$] We present a unified review of trustworthiness in intelligent education from five complementary perspectives: safety and privacy, robustness, fairness, explainability, and sustainability, while categorizing and summarizing the research approaches within each.

    \item[\small$\bullet$] We conduct a detailed discussion and summary of the shortcomings and challenges under current trustworthy intelligent education research.

    \item[\small$\bullet$] We outline and envision the future development directions of trustworthy intelligent education, including challenge-driven approaches and frontier research areas, to provide insights and guidance.

\end{itemize}

%% file: 3-intel_edu_tasks.tex
\section{Intelligent Education: Task Landscape}

Intelligent education~\cite{AIED} encompasses a wide range of learning-centered tasks that aim to support adaptive, personalized, and data-driven teaching and learning processes. Rather than viewing intelligent education as a collection of isolated techniques, a more meaningful perspective is to frame it around the core educational objectives that AI-driven systems are designed to serve. In this survey, we organize intelligent education into five representative task groups according to their roles in the educational pipeline, spanning from learner state modeling to instructional support. Figure~2 provides an overview of the proposed task taxonomy.

\subsection{Learner Ability Assessment}

Learner ability assessment is to infer learners’ potential knowledge states and learning abilities based on their observed interactions with educational systems. In this vein, \textbf{\textit{Cognitive Diagnosis}}~\cite{wang2024survey} targets the characterization of learners' proficiency in fine-granularity knowledge components as well as in interpretable and multidimensional competency profiles in which typically local stability is assumed on learners’ competencies. Distinguishing from the static nature of cognitive diagnosis, \textbf{\textit{Knowledge Tracing}}~\cite{DKT} is developed to track learners’ knowledge states evolving over time, which mandates an explicit learning modeling that depicts the progress of ability level in terms of both knowledge acquisition and concept forgetting. In addition, \textbf{\textit{Computerized Adaptive Testing}}~\cite{CAT_Survey} aims to continuously and accurately estimate test-takers' abilities in interactive testing environments by dynamically selecting adaptive learning items at the lowest cost. While these sub-tasks distinctly depend on the underlying modelling assumptions and are operational in distinct settings, they all have the same common goal of faithfully reflecting learner knowledge levels according to observed learning behaviours, and thus solidify learner ability assessment as a cornerstone component in the intelligent education field.

% figure 
\begin{figure*}[!t] %H为当前位置，!htb为忽略美学标准，htbp为浮动图形
	\centering %图片居中
        \vspace{-2mm}
	\includegraphics[scale=0.127]{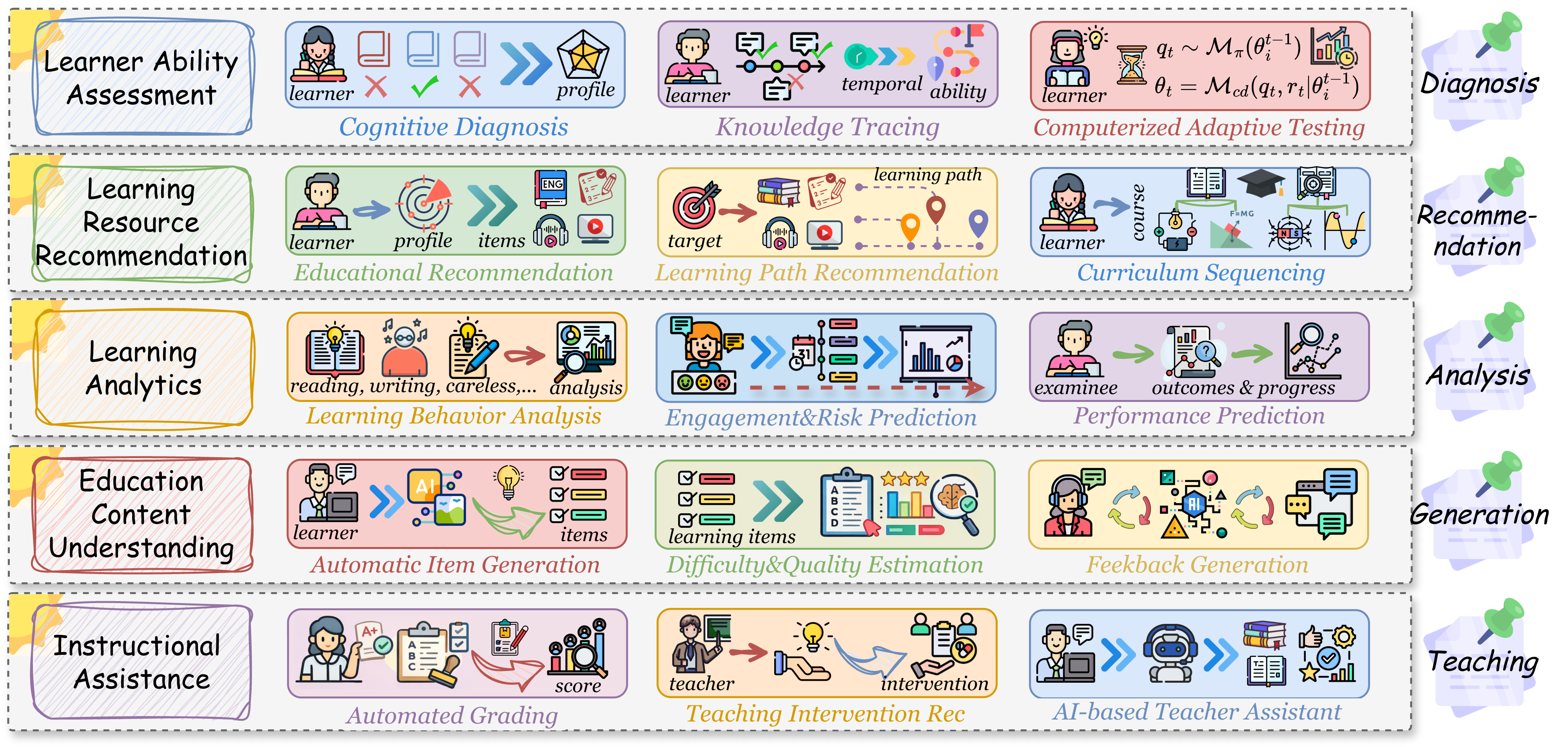}
    \vspace{-5mm}

    \caption{An overview of intelligent education task categories. Five high-level task categories are considered, including learner ability assessment, learning resource recommendation, learning analytics, educational content understanding, and instructional assistance.}

        % \vspace{-5mm}
        \label{fig.framework} %用于文内引用的标签; 必须加在caption之后
        \vspace{-5mm}
\end{figure*}

\subsection{Learning Resource Recommendation}

Learning resource recommendation aims in selecting and organizing suitable learning materials to assist learners in achieving their personalized, effective learning. In this category, \textbf{\textit{Educational Recommendation}}~\cite{Exercise_Rec_1} makes effort to recommend appropriate learning resource such as course, exercise and knowledge concepts based on the preference, ability and learning needs of learners. Besides suggesting resources, \textbf{\textit{Learning Path Recommendation}}~\cite{LPR_Graph} concerns in generating a coherent learning sequence that helped learners follow an ordered learning trajectory to progress knowledge-wise and ability-wise. At a more pedagogical-oriented level, \textbf{\textit{Curriculum Sequencing}}~\cite{Curriculum_Seq_Evo} explores the global arrangement of units and their dependencies to create structured learning plans across higher instructional levels. These sub-tasks together form a natural workflow that starts from understanding learners’ abilities and states, then moves to recommending and organizing suitable learning resources, and ultimately supports continuous learning improvement.

\subsection{Learning Analytics}

Learning analytics aims to derive insights into learning processes, behaviors, and outcomes through the systematic analysis of learners’ interaction data. At a finer analytical level, \textbf{\textit{Learning Behavior Analysis}}~\cite{Behavior_Analysis} examines learners’ interaction patterns, learning strategies, and behavioral traces to reveal how learning activities unfold over time. Building upon these behavioral signals, \textbf{\textit{Engagement and Risk Prediction}}~\cite{Engagement_Pre} focuses on identifying learners’ engagement levels and potential learning risks, such as disengagement or dropout, thereby enabling early detection and timely intervention. In addition, \textbf{\textit{Performance Prediction}}~\cite{Performance_Pre} evaluates learners’ learning outcomes and progress by analyzing assessment results and activity records, supporting data-driven assessment of learning effectiveness. Together, these sub-tasks provide a comprehensive, data-driven understanding of learning processes and outcomes, positioning learning analytics as a key mechanism for monitoring and improving intelligent education systems.

% figure 
\begin{figure*}[!t] %H为当前位置，!htb为忽略美学标准，htbp为浮动图形
	\centering %图片居中
        \vspace{-2mm}
	\includegraphics[scale=0.135]{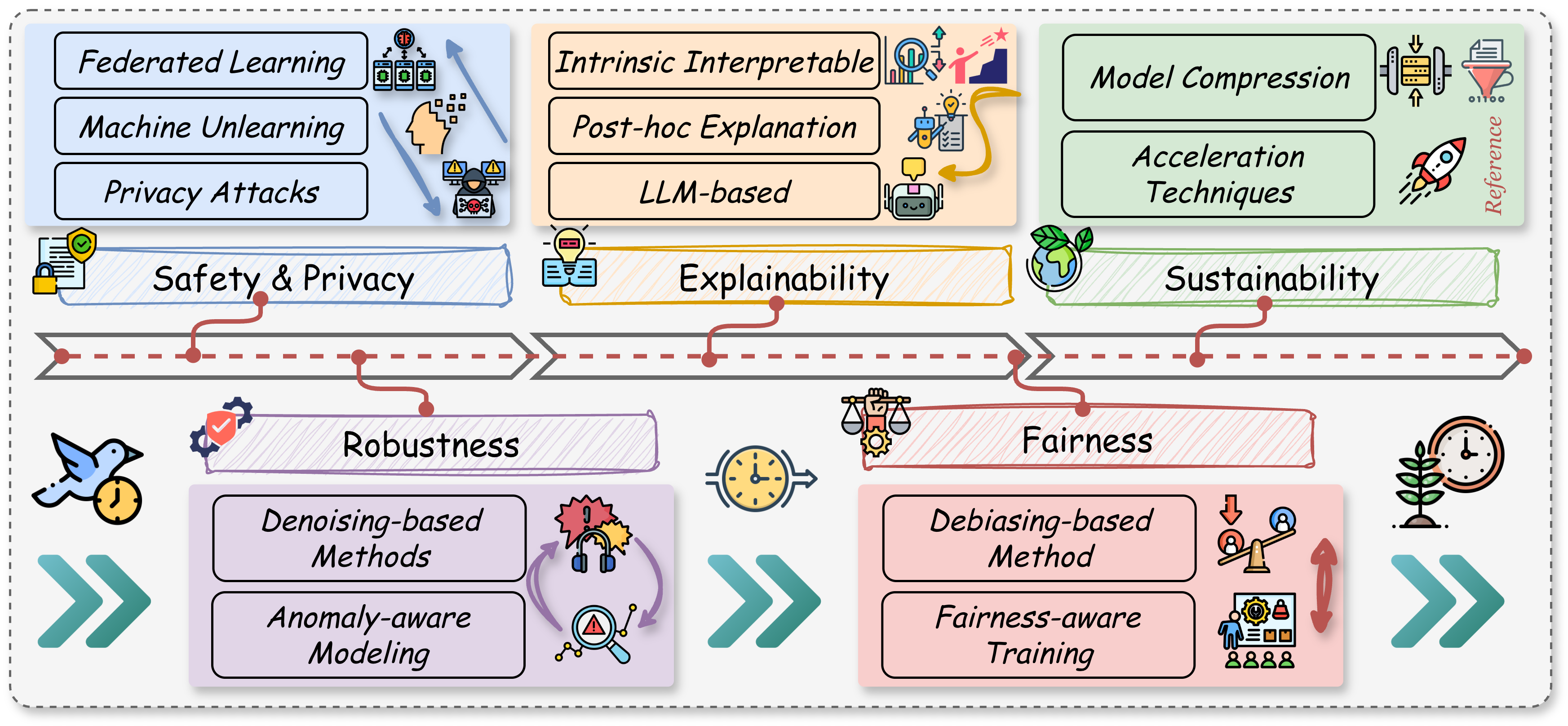} %插入图片，[]中设置图片大小，{}中是图片文件名

    \vspace{-4mm}
    \caption{An overview of research categories for trustworthy intelligent education. The categories cover five trust-related aspects and are organized to reflect an desirable trajectory, from safety\&privacy and robustness to explainability, fairness, and ultimately sustainability.}

        % \vspace{-5mm}
        \label{fig.framework} %用于文内引用的标签; 必须加在caption之后
        \vspace{-5mm}
\end{figure*}

\subsection{Education Content Understanding}

Educational content understanding is concerned with analyzing and modeling educational materials to support effective teaching and learning. One major line of work, referred to as \textbf{\textit{Automatic Item Generation}}~\cite{Auto_Item_Gen} addresses the increasing demand for large-scale and continuous assessments by automatically producing assessment items from structured inputs, thereby alleviating the cost, scalability, and item exposure issues of traditional item development. In contrast to direct generation, \textbf{\textit{Difficulty and Quality Estimation}}~\cite{Diff_Estimation_Survey} involves systematically evaluating educational materials, such as questions, based on existing learning content to determine their difficulty levels and pedagogical quality. This enables systems or instructors to select content that matches learners' abilities during the learning process. In addition, \textbf{\textit{Feedback Generation}}~\cite{Feedback_Gen} provides automated feedback based on learners’ interactive responses during practice, such as summarizing key knowledge concepts or offering explanations for answers, typically in textual form to support understanding and instructional reference. These tasks jointly cover the generation, evaluation, and instructional use of ~ content, forming the foundation for content-centered intelligence in~education.

\subsection{Instructional Assistance}

Different from the aforementioned tasks, which primarily involve facilitating exploration at the student level to promote adaptive learning within the educational process, instructional assistance provides support for teaching activities and instructional decision-making at the teacher level. \textbf{\textit{Automated Grading}}~\cite{Automated_Grading} addresses the growing burden of assessing large-scale student submissions by automatically assigning grades and output feedback in a timely and consistent manner, supporting both formative and summative assessment settings while reducing instructors’ grading workload. Beyond grading, \textbf{\textit{Teaching Intervention Recommendation}}~\cite{Teaching_Intervention} seeks to suggest prompt instructional actions or interventions based on students’ states and learning contexts, assisting educators in addressing learning difficulties and promoting progress. In addition, \textbf{\textit{AI-based Teacher Assistant}}~\cite{Teaching_Assistant} aims to providing integrated support for instructors by facilitating tasks such as instructional planning, classroom management, and pedagogical content delivery. Collectively, these tasks form a teacher-facing support layer that complements learner-centered intelligence by enabling efficient assessment, timely intervention, and informed instructional decision-making.

%% file: 4-trustworthy_IE.tex
\section{Trustworthiness in Intelligent Education}

In this section, we refer to the ethical guidelines proposed by the European Union for advancing trustworthy artificial intelligence, which stipulate that a trustworthy AI system should adhere to certain ethical principles~\cite{smuha2019eu}. We categorize and discuss trustworthiness issues and research within existing intelligent education studies primarily across five critical dimensions: \textit{Safety \& Privacy}, \textit{Robustness}, \textit{Fairness}, \textit{Explainability}, and \textit{Sustainability}, as shown in Figure~3.

\subsection{Safety \& Privacy}

Safety and privacy are consistently critical concerns in the intelligent education field. Unlike traditional educational scenarios, online learning platforms typically manage massive amounts of sensitive learner data, often involving long-term behavioral records and vulnerable user attributes, such as students' health conditions and family-related privacy information. Therefore, existing research primarily focuses on how to ensure system safety and data privacy while effectively implementing intelligent educational tasks, mainly from the perspective of adversarial attack and defense modeling.

\paragraph{Federated Learning.} One of the most common approaches to achieve privacy preservation is through the use of federated learning. Its core principle entails keeping the learning interaction data locally on client devices while only sharing updated parameters on the server side, thereby avoiding the data leakage risks associated with traditional centralized full-scale model training. FDKT~\cite{FDKT} proposes a federated knowledge tracing framework to address data scarcity, heterogeneity, and incomparability across data silos by introducing data-quality-aware local training and hierarchical model aggregation. FedCD~\cite{FedCD} introduces a fairness-aware federated cognitive diagnosis approach, designing a parameter decoupling-based personalization strategy to achieve accurate and fair student diagnosis across heterogeneous clients. Additionally, FLSD~\cite{FLSD} explores the performance of different federated methods~(FedAvg and FedProx) on the more privacy-sensitive student dropout prediction task while demonstrating the role of post-hoc AI techniques in federated contexts.

\paragraph{Machine Unlearning.} Recent advances in educational privacy-preserving have brought increased attention to the concept of machine unlearning, which refers to the process of enabling models to selectively and effectively remove previously learned information. PrivacyCD~\cite{PrivacyCD} investigates the data unlearning in cognitive diagnosis models and proposes a hierarchical importance-guided forgetting algorithm, which leverages layer-wise parameter importance to achieve efficient, precise, and utility-preserving unlearning. This algorithm provides a viable solution strategy for CD models to support the ``right to be forgotten" requirement. MU4RAAI~\cite{MU4RAAI} conducts the first structured study of machine unlearning in educational AI and proposes a reference architecture positioning MU as an effective strategy for building ethical and maintainable educational systems.

\paragraph{Privacy Attacks.} 

Beyond defense modeling, recent research has also investigated privacy attacks and auditing to assess privacy risks in intelligent education systems. Techniques such as membership inference attacks~(MIA) reveal the extent to which trained models may inadvertently leak information about individual learners. For example, P-MIA~\cite{P-MIA} explores the member inference attacks in cognitive diagnosis models under the realistic grey-box settings. By leveraging exposed knowledge state visualizations, the proposed profile-based MIA framework significantly outperforms black-box methods while enabling effective auditing of unlearning mechanisms.

\paragraph{Discussion} Despite these advances, ensuring safety and privacy of intelligent education remains inherently challenging. Educational data is often time-series, context-dependent, and typically non-randomly observed; these characteristics make it difficult to balance privacy protection with the need for fine-grained personalization. In addition, existing approaches usually address privacy concerns at specific stages of the learning pipeline, such as training or data storage, while providing limited guarantees for downstream decision-making processes, such as learning recommendation and instructional interventions. Furthermore, current research does not sufficiently consider or explore more complex or extreme scenarios, such as modeling privacy protection in multi-modal teaching environments that involve text, video, and audio data.

\subsection{Robustness} 

Due to the fact that real-world educational data is often noisy, incomplete, and influenced by irregular learner behavior, robustness is also a critical requirement for intelligent education systems. In contrast to regularized experimental settings, learning interactions on online platforms may involve careless responses, guessing behavior, missed attempts, or abrupt shifts in learning behavior, which can significantly undermine the reliability of model predictions. Therefore, existing robustness research primarily centers on filtering noise, detecting abnormal patterns, and stabilizing knowledge modeling under conditions of uncertainty and incomplete data.

\paragraph{Denoising-based Methods.}

A common class of methods for enhancing robustness involves explicitly modeling and mitigating noise in student learning data, typically through disentangling true learning signals from noisy or unreliable interaction observations before or during model training. For example, AdaRD~\cite{AdaRD} proposes an adaptive response denoising framework to enhance cognitive diagnosis models by introducing the generalized cross entropy-based robust training and the variance-guided response re-weight strategy. Similarly, DiffCog~\cite{DiffCog} leverages diffusion-based modeling to smooth noisy cognitive states, enabling more stable estimation of learners’ knowledge proficiency under corrupted interaction records.

\paragraph{Anomaly-aware Modeling.}

Alongside general noise, another category of research examines anomalous or atypical learner behavior that deviates from regular learning patterns, which we summarize as anomaly-aware modeling methods. They aim to identify and address these behaviors to prevent them from disproportionately affecting model predictions, thereby ensuring robustness. HD-KT~\cite{HD-KT} proposes a hybrid interaction-denoising framework to enhance robustness in knowledge tracing, explicitly detecting abnormal behavior through two detectors: one knowledge-state-guided and the other student-profile-guided. RobustKT~\cite{RobustKT} further decouples the cognitive pattern from random-error-induced behaviors to achieve more accurate and noise-resistant knowledge tracing by employing a cognitive decoupling analyzer and the decay-based attention.

\paragraph{Discussion}

The inherent heterogeneity and non-stationary of educational data, stemming from learners' motivation, behavior, and engagement levels that naturally fluctuate over time, remain a significant challenge in effectively distinguishing noise from valuable learning signals to enhance robustness. Moreover, existing approaches often rely on idealized assumptions about noise distributions or anomaly patterns, which may fail to address the complexity of real-world learning environments and may not generalize effectively across different learning entities, including learners, resources, and platforms. Meanwhile, current robustness-enhancing approaches are tightly coupled with specific tasks and models, limiting their applicability to broader educational pipelines involving multiple interconnected decisions, such as multi-source resource recommendations.

\subsection{Fairness}

Fairness has always been a fundamental concern in education, particularly in intelligent education, as algorithmic decisions may disadvantage certain student groups due to biased data, unequal exposure to resources, or disparities in consumption levels. Unlike recommendation systems, educational decisions often have long-term and cumulative effects, making unfair treatment particularly harmful to students’ development. Therefore, existing research on fairness in intelligent education mainly explores how to identify and mitigate biases in learner modeling and educational decision-making.

\paragraph{Debiasing-based Methods.}

An prominent category of work is based on debiasing methods, which explicitly model biases such as response bias and cognitive bias through causal learning and disentanglement learning to ensure fairness in cognitive modeling. For instance, CORE~\cite{CORE} proposes a counterfactual reasoning framework for knowledge tracing that identifies answer bias as a direct causal effect of the query. By reducing this effect through subtraction from the overall causal impact, it achieves unbiased prediction across multiple KT models and datasets. Similarly, PSCRF~\cite{PSCRF} proposes a path-specific causal inference framework that preserves useful diagnosis information while eliminating the shortcut effects of sensitive attributes. It incorporates attribute-decoupling and multi-factor constraints to achieve fair and accurate student proficiency prediction. Moreover, DisKT~\cite{DisKT} explores the counterfactual effects that cause cognitive biases in knowledge tracing from the perspective of knowledge disentanglement, and introduces a contradiction-aware attention mechanism to ensure effective modeling.

\paragraph{Fairness-aware Training.}

Beyond debiasing modeling, several studies explicitly incorporate fairness objectives into the model training process. These approaches typically introduce fairness-aware regularization terms or constraints to balance predictive performance across learner groups. FairCD~\cite{FairCD} theoretically reveals and explains the unfairness present in existing cognitive diagnosis models, then mitigates it by decomposing student proficiency into biased and unbiased components and performing orthogonal learning. Additionally, FairWISA~\cite{FairWISA} proposes a sensitive-attribute–free fairness framework for personalized learner modeling. It primarily introduces a min-max fairness objective to infer pseudo-sensitive information and retrain diagnosis models to eliminate unfairness.

\paragraph{Discussion}
Despite these efforts, achieving fairness in intelligent education remains challenging. First, offline and online educational data are deeply intertwined with social, economic, and contextual factors, making it difficult to define universally accepted fairness criteria. Moreover, many existing methods focus on short-term or single-step fairness, overlooking the long-term and sequential nature of educational decision-making. In addition, fairness-aware models often rely on sensitive attributes that may be incomplete, unavailable, or ethically problematic to collect in practice, limiting their applicability in real-world educational systems.

\subsection{Explainability}

Explainability is essential for intelligent education systems, as educational stakeholders—including students, teachers, and administrators—require transparent and explainable reasoning to effectively utilize algorithmic decisions. Unlike purely performance-driven domains, educational decision-making typically requires pedagogical justification, making interpretability a natural necessity rather than an optional feature. Existing approaches primarily address this through the use of inherently interpretable models, post-hoc, and leveraging large language models~(LLMs) for assistance.

\paragraph{Intrinsic Interpretable Models.}

A category of intuitive approaches involves directly utilizing and enhancing models with inherently interpretable structures to model educational tasks; these approaches typically align modular components with educational concepts. For example, KAN2CD~\cite{KAN2CD} employs Kolmogorov-Arnold network to replace and reconstruct the MLP components within neural diagnosis models, thereby significantly enhancing the interpretability of cognitive diagnosis while maintaining prediction performance. Meanwhile, XKT~\cite{XKT} incorporates multi-feature embeddings, cognitive processing networks, and MIRT-based neural prediction to deliver interpretable representations in knowledge tracing modeling.

\paragraph{Post-hoc Explanation Methods.}

Another widely adopted strategy involves generating explanations after model inference is complete, without modifying the underlying predictive model. This post-hoc approach aims to attribute predictions to influential inputs or underlying factors. For instance, GCE~\cite{GCE} proposes a genetic algorithm-based causal explainer for estimating sequence-level causal attribution in deep knowledge tracing models, overcoming the limitations of gradient-based and attention-based explanations.

\paragraph{LLM-based Explanation.}

With the rise of large language models~(LLMs), recent research has also explored using LLMs for interpretable educational prediction and recommendation. Specifically, these approaches leverage the generative and reasoning abilities of LLMs to translate model outputs into educationally meaningful interpretations. For example, EFKT~\cite{EFKT} introduces an explainable few-shot knowledge tracing model, in which an LLM-based cognition-guided framework infers student knowledge from limited data while generating natural-language explanations. Moreover, LACAT~\cite{LACAT} employs the LLM-based agent framework to augment computerized adaptive testing, achieving human-like interpretability through its summarization, reasoning, and critique modules.

\paragraph{Discussion}

Although interest in intelligen education continues to grow, its interpretability remains constrained by several factors. First, intrinsically interpretable models often sacrifice prediction performance for transparency, while post-hoc explanations may fail to faithfully reflect the model's internal reasoning. Similarly, while LLM-based explanations offer rich and intuitive expressions, underlying issues of hallucinations and pedagogical reliability persist. Furthermore, existing work has primarily focused on interpreting model outputs, with limited attention paid to explaining long-term educational decisions or their pedagogical implications.

\subsection{Sustainability}

Due to the common large-scale and long-term deployment of intelligent education platforms, models are typically required to maintain efficiency, adaptability, and resource constraints. Consequently, sustainability has recently emerged as an increasingly focal aspect of trustworthy intelligent education, including computational efficiency, maintainability, and the scalability of educational systems. Current research broadly falls into two categories in exploring sustainable intelligent education: model compression and acceleration techniques.

\paragraph{Model Compression.}

In educational modeling, models primarily consist of embedding layers and interaction layers. The former is used to map discrete features, for instance, in cognitive diagnostics, student IDs are typically mapped into more expressive dense vectors, while the latter models the complex interactions between learners and learning elements to enable prediction, such as MLPs and self-attention networks. Consequently, current model compression approaches mainly encompass model-level neural architecture search methods and parameter-level hashing strategies. Specifically, ENAS-KT~\cite{ENAS-KT} proposes an evolutionary neural architecture search framework to enhance the design of transformer-based knowledge tracing models by introducing convolutional local context modeling and an automatic feature selection operator. In contrast, HashCAT~\cite{HashCAT}, begins with efficient embedding representations, proposing a learning-to-hash-based method to generate binary representations of students and exercises, and utilizes an entropy-driven uncertainty reduction strategy to accelerate item selection in computerized adaptive testing.

\paragraph{Acceleration Techniques.}

As for acceleration techniques, these involve proposing effective approaches based on existing models to accelerate model training or inference, thereby achieving efficient educational modeling and decision-making. For example, SECAT~\cite{SECAT} proposes an search-efficient computerized adaptive testing framework, designing a student-aware item indexing technique and multi-round logarithmic search strategy to accelerate question selection during assessment. Notably, it reduces the time complexity from $O(N)$ to $O(logN)$, achieving over 200x speed up with negligible accuracy degradation.

\paragraph{Discussion}

Despite progress, sustainability in intelligent education still faces enduring challenges. Specifically, many effective models are optimized for specific tasks or datasets, limiting their generalizability across different educational scenarios. In addition, efficiency-oriented methods may overlook long-term maintainability issues particularly in online systems experiencing explosive data growth, such as model updating, concept drift, and evolving curricula. Notably, balancing performance, efficiency, and adaptability remains an open challenge for sustainable intelligent education systems.

%% file: 5-future_directions.tex
\section{Future Directions}
\subsection{Challenge-driven Directions}

\paragraph{Holistic trustworthiness cross pipelines.}

Existing research on trustworthy intelligent education mainly addresses trustworthy issues for specific tasks in isolation, neglecting dependencies between educational pipelines. Future research can explore comprehensive trustworthy frameworks to achieve trustworthiness assurance across the entire real educational process, from data interaction to cognitive modeling, and to recommendation, decision-making, and instruction.

\paragraph{Long-term and sequential trust modeling.} 

Trust in education is fundamentally a long-term and continuous endeavor, yet existing methods treat it merely as a simple decision point. Future work could explore modeling this long-term trustworthiness problem, investigating its temporal evolution-for example, fairness in knowledge tracing may evolve alongside the development of student populations and capabilities.

\paragraph{Trust under realistic and imperfect data.} Existing trustworthy intelligent education methods often rely on simplified assumptions about data quality and distribution that may not accurately reflect the real learning environments. Future research should focus on trust-aware modeling under realistic and imperfect data conditions, explicitly accounting for noise, missingness, and distributional shifts in educational data.

\subsection{Emerging Research Frontiers}

\paragraph{Multi-modal trustworthy education.}

Beyond structured interaction records, real-world adaptive learning environments increasingly engage with rich multi-modal features such as text, image, speech, video, and behavioral traces, while existing trustworthy education studies remain limited to single-modal data. This modality gap not only constrains the modeling of authentic learning behaviors but also introduces new risks related to cross-modal bias, privacy leakage, and robustness degradation. Future research should explore multi-modal trustworthy education modeling that jointly addresses privacy, robustness, fairness, explainability, and sustainability across heterogeneous modalities and their interactions, enabling more reliable and realistic educational systems.

\paragraph{LLM-enabled trustworthy education.}

With the rapid adoption of large language models in educational applications, LLMs have demonstrated strong potential in tasks such as content generation and instructional support, yet their trustworthiness remains insufficiently explored. In particular, issues such as hallucination, uncontrolled bias, and unstable reasoning raise concerns when LLMs are deployed in high-stakes educational settings. Future work may focus on systematically evaluating and enhancing the safety, reliability, and other aspects of trustworthiness in LLM-driven intelligent educational systems, ensuring their alignment with pedagogical objectives and educational values.

\paragraph{Multi-agent educational modeling.}

As intelligent education evolves toward system-level analysis, multi-agent frameworks have been increasingly explored to simulate complex educational environments involving students, teachers, and institutions. Such simulation-based paradigms offer new opportunities for policy evaluation, intervention testing, and large-scale educational analysis without directly affecting real learners. Future research should explore trust-related issues in agent-based learning simulation, such as the reliability of student behavior modeling, the robustness of simulated policies, and the ethical risks of deploying decision-making strategies derived from synthetic educational environments.

\paragraph{Trust-aware evaluation and benchmarking.}

From an evaluation perspective, current assessments of intelligent education methods still mainly emphasize predictive accuracy, providing limited insight into trust-related properties. This evaluation gap makes it difficult to compare methods beyond performance or to understand their behavior under realistic constraints. Future research should develop standardized benchmarks and evaluation protocols that explicitly measure safety, robustness, fairness, explainability, and sustainability in educational settings, supporting more comprehensive and trustworthy model assessment.

%% file: 6-conclusion.tex
\section{Conclusion}

This survey presented a systematic and structured survey of trustworthy intelligent education. We first clarified the scope of intelligent education and emphasize the importance of trustworthiness given the sensitivity and decision-critical nature of educational scenarios. We then identified five task categories that characterize the core functionalities of intelligent education systems. Based on this task landscape, we examineed existing studies from five trustworthiness perspectives, namely safety and privacy, robustness, fairness, explainability, and sustainability. Through this structured organization, the survey revealed the fragmented nature of current research and highlights common limitations and gaps. Finally, by summarizing open challenges and future directions, this work provided a consolidated reference for understanding and advancing trustworthy intelligent education.